\documentclass[10pt,twoside]{article}
\usepackage{amssymb}
\usepackage{amsmath}
\usepackage[dvips]{graphicx}
\usepackage[latin1]{inputenc}
\usepackage{pifont}

\providecommand{\doint}{\oint}

\begin{document}

\title{The role of the virtual work in Faraday's law.}
\author{Rodolfo A. Diaz\thanks{%
radiazs@unal.edu.co}, William J. Herrera\thanks{%
jherreraw@unal.edu.co}, Shirley Gomez\thanks{%
sgomezp@unal.edu.co} \\
%EndAName
Departamento de Física. Universidad Nacional de Colombia. Bogotá, Colombia.}
\date{}
\maketitle

\begin{abstract}
In the context of Faraday's induction law, we show that the concept of
virtual work provides another point of view to clarify the nature of the
induced electric field, as well as the fact that the integral over a closed
path of the induced electric field is not the work performed by a unit
charge. The usefulness of the concept of virtual conservativity is
discussed. Further, we study the relation between the electromotive force
and the real work done by an external agent to keep a circuit at constant
velocity. From this discussion it is observed that magnetic forces can be
non-conservative from the mathematical point of view, but can be treated as
conservative for all practical purposes.


\end{abstract}

\section{Introduction}

The central nature of the electrostatic field leads to its conservative
character, and its field lines start and finish on charges. Faraday's
experiments revealed that electric fields with non-null circulation can be
created (in absence of net charge) by time-varying magnetic fluxes. These
electric fields have closed field lines, so that they are non-conservative.
Faraday's law has a crucial role as a theoretical tool \cite{Taussig}-\cite%
{Purcell} as well as in applications \cite{saslow}. Therefore, we encounter
several points of view to explain Faraday's Law and the nature of the
induced electric field \cite{Taussig}-\cite{Dragan2}. Nevertheless, it is
not usual to distinguish between the non-conservative behavior of induced
fields\footnote{%
In this document, the concept of induced electric field will be used with
its meaning in the context of Faraday's law.} with respect to the
non-conservative behavior of fields that come from charges in motion. In the
same way, no distinction is usually done concerning closed integrals of the
field defined in a fixed time, with the real work on a closed path done over
a unit charge. In this paper, we analyze the role of the virtual work as an
alternative tool to understand Faraday's induction law and to find the
nature of the induced electric field, taking into account that the integral
over a closed path of such a field is not the work performed by a unit
charge. In addition, we study the relation between the electromotive force
and the real work done by an external agent to keep a circuit at constant
velocity. The latter relation leads us to discuss some subtleties concerning
the conservative nature of the magnetic forces.

\section{Faraday's induction law}

\begin{figure}[t]
\begin{center}
\includegraphics[scale=0.7]{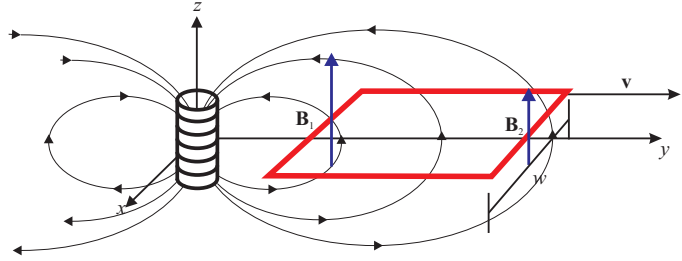}
\end{center}
\caption{A rectangular loop moving at constant velocity $\mathbf{v}$, in the
field generated by a long conducting wire. $\mathbf{B}_{1}$ is the field in
the segment of the loop that is closer to the wire, while $\mathbf{B}_{2}$
is the field in the farthest segment, $w$ is the width of the loop.}
\label{fig1}
\end{figure}

In order to initiate the discussion, we present briefly the customary
approach on Faraday's law. Consider a charge $q$ in a rectangular loop of
wire that moves at constant velocity $\mathbf{v}$, in a non-uniform
stationary magnetic field $\mathbf{B}$, (e.g. produced by a long conducting
wire, as shown in Fig. \ref{fig1}). We shall calculate the line integral of $%
\mathbf{f}\cdot d\mathbf{r}$ over a closed path in the rectangular loop,
where $\mathbf{f\ }$is the force due to the magnetic force on the charge $q$%
. In the segment closer to the wire, charges tend to circulate in
counter-clockwise sense (observed from above), while in the segment farther
to the wire, charges tend to circulate in the clockwise sense as observed
from above. Since $\mathbf{B}$ is more intense in the closer segment, net
circulation goes in counter-clockwise sense. As for the segments that are
parallel to $\mathbf{v}$, the displacement of $q$ and the force on it are
perpendicular, so they give no contribution to $\mathbf{f}\cdot d\mathbf{r}$%
. Taking the contribution from the segments perpendicular to $\mathbf{v}$,$%
\mathbf{\ }$the so-called electromotive force yields.

\begin{equation}
\varepsilon =\frac{1}{q}\oint \mathbf{f}\cdot d\mathbf{r}=vw\left( \mathbf{B}%
_{1}-\mathbf{B}_{2}\right)  \label{ec1}
\end{equation}%
where $\mathbf{B}_{1}$,\ $\mathbf{B}_{2}$ and $w$ are defined in Fig. \ref%
{fig1}. In textbooks, it is customary to argue that when the charge moves
very rapidly, the integral (\ref{ec1}) represents the work realized over $q$%
\ \cite{Purcell}. We take a textual fragment from Ref. \cite{Purcell} Sec.
7.3

\emph{\textquotedblleft If we imagine a charge }$q$ \emph{to move all around
the loop, in a time short enough so that the position of the loop has not
changed appreciably, then Eq. (\ref{ec1}) gives the work done by the force }$%
\mathbf{f}$\emph{\ per unit charge\textquotedblright .}

Nevertheless, owing to the form of the Lorentz force $q\mathbf{v}\times 
\mathbf{B}$, the magnetic field cannot do work regardless the velocity of
the charge. The contradiction comes from the fact that the line integral in
Eq. \emph{(\ref{ec1})}$,\ $is carried out at a \textbf{fixed time}.
Therefore, it does not correspond to a real work, since a real work must
consider time-evolution \cite{Griffiths, mosca}. In our subsequent
developments, we shall show the following (a) Equation (\ref{ec1}) can also
be interpreted as a virtual work done by the magnetic force $\mathbf{f}$ on
the charge $q$. (b) In some cases $\varepsilon $ coincides with the real
work done by an external agent to keep a circuit at constant velocity%
\footnote{%
This item was shown by Refs. \cite{Griffiths, mosca} but with some mistakes
in the procedure. Indeed, the correction of the procedure led us to some
subtleties about the conservative nature of the magnetic force.}. (c)
Magnetic forces are not conservative from the mathematical point of view but
they can be considered as conservative for all practical purposes.

By now, we come back to the problem of the rectangular loop. Let us consider
the flux due to the magnetic field through the rectangular loop. It is
independent of the form of the surface limited by the loop because $\nabla
\cdot \mathbf{B}=0$. The change of flux along the time $dt$ gives

\begin{equation}
d\Phi =-vw\left( \mathbf{B}_{1}-\mathbf{B}_{2}\right) dt  \label{ec2}
\end{equation}%
combining Eqs. (\ref{ec1}, \ref{ec2}) we find%
\begin{equation}
\varepsilon =-\frac{d\Phi }{dt}  \label{ec3}
\end{equation}%
and this relation is valid for an arbitrary form and velocity of the loop 
\cite{Purcell, Lopez-pique}. Moreover, for an observer $F^{\prime }$ that
moves with the loop, the electric and magnetic fields are$\ \mathbf{E}%
^{\prime }$ and $\mathbf{B}^{\prime }$ respectively. Since for $F^{\prime }$
the loop is at rest, the electromotive force (emf)\ comes exclusively from
the electric field, and we find%
\begin{equation}
\varepsilon ^{\prime }=\doint \mathbf{E}^{\prime }\cdot d\mathbf{r}^{\prime
}=-\frac{d\Phi ^{\prime }}{dt^{\prime }}  \label{ec4}
\end{equation}

When we extrapolate the previous equations to an arbitrary closed path $C$
which is stationary with respect to an inertial frame $F$,\ we obtain
Faraday's induction law

\begin{equation}
\varepsilon =\oint\limits_{C}\mathbf{E}(\mathbf{r},t)\cdot d\mathbf{r}=-%
\frac{d}{dt}\int_{S}\mathbf{B}(\mathbf{r},t)\cdot d\mathbf{S}  \label{ec5}
\end{equation}%
where $S$ is a surface\ which has $C$ as its boundary. It is essential to
emphasize that the closed line integral for the electric field in Eq. (\ref%
{ec5}), is performed for a \textbf{fixed value of\ time.}

\section{Virtual work}

In order to analyze the role of the virtual work \cite{Srivastava, Healy} in
Faraday's law, it is important to point out the nature of the electromotive
force (emf). Under the conditions of such a law, the emf owes only to the
electric field, because the closed path is assumed stationary with respect
to the system of reference that measures the flux. A second aspect is that
the emf is calculated as an integral over a closed path in which each
differential element is calculated at the same value of time $t$. For
example, two contributions in two different segments of the path $\mathbf{E}%
(x_{1},y_{1},z_{1},t)d\mathbf{r}_{1}$ and $\mathbf{E}(x_{2},y_{2},z_{2},t)d%
\mathbf{r}_{2}$ are evaluated at the same time $t$. It implies that such an
integral is not the work that a point unit charge would do to close the
circuit, for if the field is a function of time, a differential of it must
be calculated using the value of the field in the space-time point where the
particle is located, two small contributions for this real work must be of
the form $\mathbf{E}(x_{1},y_{1},z_{1},t_{1})~d\mathbf{r}_{1}$ and $\mathbf{E%
}(x_{2},y_{2},z_{2},t_{2})~d\mathbf{r}_{2}$.

On the other hand, to calculate the work done by the magnetic force in Eq. (%
\ref{ec1}), it was assumed that the line integral is realized at a fixed
instant of time. We call this a \textbf{virtual work}, which is the work
done on a charge when time is considered fixed along the path of
integration. In the virtual displacement defined here, the only constraint
is that the particle must follows the prescribed path $C$. Notice that it
does not correspond to the real work that would be done on a charge, since
real works must consider time evolution. In that sense, we can say that
magnetic forces cannot do real work, but they can do virtual work. According
to this discussion, $d\mathbf{r}$ in Eq. (\ref{ec1}) is a virtual
displacement.

Further, it is well-known that in the general case, the induced field comes
from the relativistic transformation of the fields from the system $F$ to
the system $F^{\prime }$ in which the loop is stationary, where $F^{\prime
}\ $is connected with $F$ by a boost. In $F^{\prime }$ the contribution to
the closed line integral comes exclusively from the electric field, and we
see that this closed line integral is not zero. Now, since this field comes
from a relativistic transformation, it is not generated by charges.

The concept of virtual work permits us to see the same facts under another
point of view. It is clear that the electric fields generated by charges in
motion are in general non-conservative because of the explicit
time-dependence. Nevertheless, these fields are \textbf{virtually
conservative}. To show this, observe that in a virtual process, owing to the
lack of time evolution, the configuration of charges that are the sources of
the field are fixed as we travel over the path to calculate the line
integral. Therefore, in a virtual process an electric field coming from
charges behaves as an electrostatics field, and therefore it is virtually
conservative\footnote{%
Of course there is a (instantaneous) magnetic field generated by the charges
in motion. But this emf arises from the electric field only.}. By contrast,
the induced field involved in Faraday's law is not virtually conservative,
since the virtual work (emf) over a closed path is non-zero. It implies that
such a field cannot come from charges in motion. In that sense, the closed
path integral in Faraday's law is not the correct quantity to evaluate real
conservativity, except under certain approximations.

There is still another way to see the difference between fields coming from
charges, and induced fields. For fields generated by charges, the condition $%
\nabla \times \mathbf{E}_{charge}=0$ is satisfied in the whole space and at
any given time, while for induced fields $\nabla \times \mathbf{E}%
_{induc}\neq 0$ in at least some points of space-time. The condition $\nabla
\times \mathbf{E}_{charge}=0$ is necessary and sufficient for conservativity
in electrostatics. Notwithstanding, for charges in motion, in which $\mathbf{%
E}$ is a function of time, the nullity of the curl only guarantees that $%
\mathbf{E}=-\nabla \phi (\mathbf{r},t)$ which in turn implies that a virtual
work yields 
\begin{equation*}
q\int_{r_{A}}^{r_{B}}\mathbf{E}\cdot d\mathbf{r}=q\left[ \phi (\mathbf{r}%
_{B},t)-\phi (\mathbf{r}_{A},t)\right]
\end{equation*}%
and it is independent of the path since time is fixed in the process.
However, in a real trajectory the work can depend on the path because of the
time evolution. The conclusion is that $\nabla \times \mathbf{E}=0$ in the
whole space at any time, is equivalent to \textbf{virtual conservativity}
(i.e. conservativity for virtual works). Of course, in the electrostatic
case virtual works coincide with real works so that in this case, virtual
conservativity is equivalent to real conservativity.

Despite the induced field $\mathbf{E}$ is not virtually conservative, nor is
originated in charges, the force undergone by a charge immersed in this
field takes the form $\mathbf{F}=q\mathbf{E}$, which is the expression found
by fields generated by charges. So although $\mathbf{E}_{charges}$ and $%
\mathbf{E}_{induc}$ come from different sources, they accomplish the same
local property with which electric field was initially defined. From the
previous discussion the total electric field can be decomposed as

\begin{equation}
\mathbf{E=E}_{charge}+\mathbf{E}_{induc},  \label{ec6}
\end{equation}%
where 
\begin{eqnarray*}
\nabla \times \mathbf{E}_{charge} &=&0\ \ \ ;\ \ \ \nabla \times \mathbf{E}%
_{induc}=-\frac{\partial \mathbf{B}}{\partial t} \\
\nabla \cdot \mathbf{E}_{charge} &=&\frac{\rho }{\varepsilon _{0}}\ \ ;\ \
\nabla \cdot \mathbf{E}_{induc}=0
\end{eqnarray*}%
So that $\mathbf{E}_{charge}$ is irrotational (i.e. virtually conservative),
and its field lines start and finish on charges. By contrast, $\mathbf{E}%
_{induc}$ is solenoidal (charges are not its sources) and it is not
virtually conservative.

\section{Electromotive force as a real work}

\begin{figure}[tbh]
\begin{center}
\includegraphics[width=7.2cm]{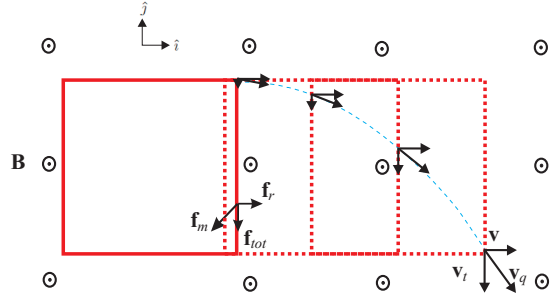}
\end{center}
\caption{A U-shape conductor with a conducting road sliding on it at
constant velocity, by the action of an external agent. The system is
immersed in a uniform magnetic field going out of the paper. In our picture,
we have assumed $q>0$, but all equations are valid for an arbitrary value
and sign of $q$.}
\label{fig:cuadrado}
\end{figure}

We have related the emf with the concept of virtual work. In this section,
we shall see that in some cases the emf can also be related with real works.
To show this, let us consider a U-conductor with a conducting rod sliding on
it with constant velocity $\mathbf{v}$. The system is immersed in a uniform
magnetic field $\mathbf{B}=B\widehat{\mathbf{k}}$ going out of the paper
(see Fig. \ref{fig:cuadrado}). There are two forces acting on the charge $q$%
, the magnetic force\textbf{\ }$\mathbf{f}_{m}$ due to $\mathbf{B}$, and $%
\mathbf{f}_{r}$ which is a \textquotedblleft drag\textquotedblright\ force
that is provided for an external agent (to keep the velocity of the rod
constant), and transmitted to the charge by the structure of the wire%
\footnote{%
Microscopically, this drag force is due to a Hall effect inside the wire,
see Ref.\cite{mosca}.}. Hence, the total force $\mathbf{f}_{tot}\ $on the
charge $q$,\ yields%
\begin{equation}
\mathbf{f}_{tot}=\mathbf{f}_{r}+\mathbf{f}_{m}  \label{sum force}
\end{equation}%
if we denote $\mathbf{v}_{t}$ the tangential velocity of the charge, the
real velocity of the charge is 
\begin{equation*}
\mathbf{v}_{q}=\mathbf{v}+\mathbf{v}_{t}=v\mathbf{\hat{\imath}}+v_{t}\mathbf{%
\hat{\jmath}}
\end{equation*}%
in a similar way a real infinitesimal displacement of the charge\ $d\mathbf{L%
}_{q}$ is given by%
\begin{equation*}
d\mathbf{L}_{q}=d\mathbf{L}+d\mathbf{L}_{t}=\mathbf{\hat{\imath}~}dL+\mathbf{%
\hat{\jmath}~}dL_{t}
\end{equation*}%
The $x-$component of the total force on $q$ is zero because $\mathbf{v}$ is
constant. Hence

\begin{eqnarray}
\mathbf{f}_{r}\cdot \mathbf{\hat{\imath}} &=&-\mathbf{f}_{m}\cdot \mathbf{%
\hat{\imath}}=-q\left[ \left( \mathbf{v}+\mathbf{v}_{t}\right) \times 
\mathbf{B}\right] \cdot \mathbf{\hat{\imath}}  \notag \\
\mathbf{f}_{r}\cdot \mathbf{\hat{\imath}} &=&-qv_{t}B=f_{r}  \label{fr as vt}
\end{eqnarray}%
Now, the total force along the $y-$direction is given by the $y-$component
of the magnetic force

\begin{equation}
f_{tot}=\mathbf{f}_{tot}\cdot \mathbf{\hat{\jmath}}=\mathbf{f}_{m}\cdot 
\mathbf{\hat{\jmath}}=-qvB  \label{ftot is const}
\end{equation}%
hence the net force on $q$ is constant and in the $y-$direction. Therefore,
the trajectory of the charge is a parabola (and not a straight line as
assumed in Ref. \cite{mosca}). Calculating the work made by $\mathbf{f}_{r}$%
, we obtain

\begin{equation*}
dW_{r}=\mathbf{f}_{r}\cdot d\mathbf{L}_{q}=f_{r}\mathbf{\hat{\imath}}\cdot
\left( \mathbf{\hat{\imath}~}dL+\mathbf{\hat{\jmath}~}dL_{t}\right)
=f_{r}~dL=-qBv_{t}\left( t\right) ~v~dt
\end{equation*}%
Integrating for a charge starting at the top of the rod and ending at the
bottom of it, we have%
\begin{equation*}
W_{r}=-qBv\int_{t_{0}}^{t_{f}}v_{t}\left( t\right) ~dt=-qBv\int_{\mathbf{r}%
_{0}}^{\mathbf{r}_{f}}dL_{t}
\end{equation*}%
if $q>0$, the charge goes downward from top to bottom, hence 
\begin{equation*}
\int_{\mathbf{r}_{0}}^{\mathbf{r}_{f}}dL_{t}=\int_{w}^{0}dL_{t}=-w
\end{equation*}%
if $q<0$ this integral is $w$. Therefore

\begin{equation*}
W_{r}=\left\vert q\right\vert Bvw
\end{equation*}%
Now, since the magnetic force does not work, $W_{r}$ must coincide with the
work done by the total force $\mathbf{f}_{tot}$ as can be checked easily

\begin{equation*}
W_{f_{tot}}=\int_{\mathbf{r}_{0}}^{\mathbf{r}_{f}}\mathbf{f}_{tot}\cdot d%
\mathbf{L}_{q}=-qvB\int_{\mathbf{r}_{0}}^{\mathbf{r}_{f}}dL_{t}=\left\vert
q\right\vert Bvw
\end{equation*}%
further, we can observe that these real works coincide with the virtual work
done by the magnetic force, i.e. with the emf%
\begin{equation*}
W_{m}^{virt}=\int \mathbf{f}_{m}\cdot d\mathbf{L}_{t}=\int_{\mathbf{r}_{0}}^{%
\mathbf{r}_{f}}\left( \mathbf{f}_{m}\cdot \mathbf{\hat{\jmath}}\right)
~dL_{t}=-qvB\int_{\mathbf{r}_{0}}^{\mathbf{r}_{f}}dL_{t}=\left\vert
q\right\vert Bvw
\end{equation*}%
this result coincides with the one found in Ref. \cite{mosca}. It is
understandable since the total force is conservative, so despite Ref. \cite%
{mosca} is using a wrong trajectory, the length covered along the $y-$axis
is the same. However, the fact that we have a uniform acceleration instead
of a uniform velocity brings an apparent paradox as we discuss below.

Since the motion along the $y-$axis is accelerated, the speed $v_{t}$
depends on time. In turn it implies that $\mathbf{f}_{r}$ depends explicitly
on time according with Eq. (\ref{fr as vt}). Further Eq. (\ref{ftot is const}%
) shows that the total force $\mathbf{f}_{tot}$ is constant. Consequently $%
\mathbf{f}_{m}$ must also depend explicitly on time to cancel the
time-dependence of $\mathbf{f}_{r}$ in Eq. (\ref{sum force}). But forces
that depends explicitly on time are non-conservative, so $\mathbf{f}_{m}$ is
a force that depends explicitly on time but never does real work!.

This paradox has to do with the initial definition of a conservative vector
field. A vector field $\mathbf{F\ }$is conservative if for \textbf{any
trajectory}, the line-integral along such a trajectory only depends on the
extreme points. In the case of magnetic forces, we only take the path traced
by \textbf{the real trajectory of the particle}, if we calculated the line
integral along other trajectories (with fixed extremes) we would obtain in
general a result dependent on the particular trajectory\footnote{%
Two identical paths define different trajectories if they travel with
different time dependence. For vector fields depending explicitly on time,
identical paths covered in different time intervals lead to different values
of the line integral.}. Furthermore, if we assume an arbitrary magnetic
field and that velocities depend on the position, the explicit calculation
of the rotational of the magnetic force gives in general a non-zero value.
This is related with the fact that there is no potential function $\phi \ $%
that depends on the position only, and that generates the magnetic force
through the relation $\mathbf{F}=-\nabla \phi $.

To solve the puzzle we simply observe that in Physics we are not concern
about arbitrary line-integrals, we only mind line-integrals over real
trajectories of the particles to calculate works. When applied to these real
trajectories, magnetic forces produce null line-integrals, line-integrals of
such forces on other trajectories are just a mathematical curiosity. We
could say that magnetic forces are not conservative from the mathematical
definition, but they can be treated as conservative for practical purposes.
Hence, it is natural to have a magnetic force that depends explicitly on
time (as it is our case), but never does a work.

\section{Conclusions}

The introduction of the concept of virtual work to analyze Faraday's
induction law permits to clarify the nature of the induced electric field,
which does not come from charges, and the fact that the integral over a
closed path of the induced electric field is not the real work that the
field would do on a unit charge along the closed path. It also allows to
understand why in certain systems of reference, the closed line integral of
the magnetic force is not zero, despite magnetic fields cannot do work.

We studied the problem of a U-conductor with a conducting rod sliding on it
with constant velocity $\mathbf{v}$, in which the system is immersed in a
uniform magnetic field. We observe that the emf in this system coincides
with the real work done by the external agent to keep constant the velocity
of the rod. In this example, the magnetic force depends explicitly on time
and thus is not conservative as a vector field in the mathematical sense. We
observe however, that magnetic forces can be treated as conservative for
practical purposes in Physics.

On the other hand, the fact that $\nabla \times \mathbf{E}=0$ for the whole
space at any time, is equivalent to virtual conservativity of $\mathbf{E}$,
that is conservativity in a process that does not involve time-evolution. In
general, it differs from real conservativity in which time evolution is
essential. Electric fields coming from charges are virtually conservative
(and sometimes conservative), while induced electric fields in the Faraday
sense are not virtually conservative. Therefore, the concept of virtual
conservativity is useful to distinguish between induced electric fields in
the Faraday sense, and electric fields generated by charges.

It is worth pointing out that the concept of virtual conservativity can be
extended to any vector field in Physics, and therefore could be useful in
Mechanics and other scenarios of Physics.

\textbf{Acknowledgments}. The authors acknowledge to the Vicerrectoría de
investigación of Universidad Nacional de Colombia, and División de
Investigación sede Bogotá (DIB), for their financial support.

\end{document}